\documentstyle[12pt,epsfig,cite]{article}
     \newlength{\dinwidth}                       
     \newlength{\dinmargin}                      
\def\lsim{\mathrel{\rlap{\lower4pt\hbox{\hskip1pt$\sim$}}
    \raise1pt\hbox{$<$}}}                
\def\gsim{\mathrel{\rlap{\lower4pt\hbox{\hskip1pt$\sim$}}
    \raise1pt\hbox{$>$}}}                

\newcommand{\av}[1]{\mbox{$ \langle #1 \rangle $}}

\newcommand{\kt}{\mbox{$k_T~$}}

\newcommand{\PI}{{\Pi}}
\renewenvironment{thebibliography}[1]
        {\begin{list}{\arabic{enumi}.}
        {\usecounter{enumi}\setlength{\parsep}{0pt}
         \setlength{\itemsep}{0pt}
         \settowidth
        {\labelwidth}{#1.}\sloppy}}{\end{list}}

\newcommand{\ZPC}[3]{Z. Phys. {\bf{#1}} ({#2}) {#3}}

\newcommand{\CPC}[3]{Comp. Phys. Comm. {\bf{#1}} ({#2}) {#3}}
\newcommand{\NP}[3]{Nucl. Phys. {\bf#1} ({#2}) {#3}}
\newcommand{\PL}[3]{Phys. Lett. {\bf#1} ({#2}) {#3}}

\begin{document}
\bibliographystyle{unsrt}

\vspace*{10mm}
\begin{center}  \begin{Large} \begin{bf}
Transverse Momentum Transfer and \\
Low \boldmath{$x$} Parton Dynamics at HERA \\
  \end{bf}  \end{Large}
  \vspace*{5mm}
  \begin{large}
E.\@ A.\@ De Wolf and P.\@ Van Mechelen\\ {\ }\\
  \end{large}
Universitaire Instelling Antwerpen,
Department of Physics \\
Universiteitsplein 1.
B-2610 Antwerpen, Belgium \\
e-mail : edewolf@uia.ua.ac.be, pierre@uia.ua.ac.be
\end{center}
\begin{quotation}
\noindent
{\bf Abstract:}
The transverse momentum transfer correlation is introduced as a sensitive probe that can
be used to discrimate between models for parton dynamics in low-$x$ deep-inelastic scattering.
Expectations for uncorrelated models and models with short-range or long-range correlations are
discussed and confronted to results obtained from the Lepto and Ariadne Monte Carlo simulation programmes.
\end{quotation}

\section{Introduction}

\begin{figure}[htp]
   \centering
   \vspace{-0.2cm}
   \epsfig{file=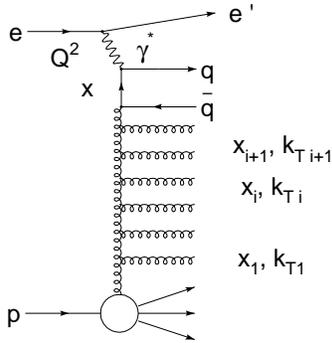,width=5cm}
   \caption{ 
      Parton evolution in the 
      ladder approximation. The longitudinal
      fractional momenta $x_i$ and transverse momenta $k_{Ti}$
      of subsequently emitted partons are indicated.   }
   \label{fig:ladder} 
\end{figure}

The successful description of the nucleon structure
function by perturbative QCD, using the DGLAP
(Dokshitzer-Gribov-Lipatov-Altarelli-Parisi) parton evolution
equations~\cite{dglap} constitutes a major success of QCD. 
However, at very small Bjorken-$x$, 
these equations are expected to become invalid. 
An alternative ansatz for the small-$x$ regime is the BFKL 
(Balitsky-Fadin-Kuraev-Lipatov) equation~\cite{bfkl1}. 
At lowest order, the BFKL and DGLAP equations resum 
the leading logarithmic $(\alpha_s\ln{1/x})^n$) or 
$(\alpha_s\ln{Q^2/Q_0^2})^n$) contributions, respectively, 
with $Q^2$ being the virtuality of the exchanged photon in
deep-inelastic neutral current $ep$ collisions. 
The leading-log DGLAP ansatz implies strong ordering 
($Q_0^2\ll {k_T}_1^2 \cdots \ll {k_T}_i^2 \ll \cdots Q^2$) 
of the transverse momenta $(k_T)_i$ in the parton cascade.
Here, $(k_T)_i$ is the transverse momentum 
(measured relative to the incident proton direction) 
of the  $i$-th emitted ``final state'' parton (see Fig.\@~\ref{fig:ladder}).
According to BFKL kinematics, the transverse momenta are no longer
strongly ordered but may be thought of 
as forming, roughly speaking, a random walk  with 
$(k_T^2)_i \sim(k_T^2)_{i+1}$.

Measurements of the hadronic final state resulting from the
initial-state parton cascade should be sensitive to the type of evolution.
For example, without strong $k_T$-ordering, more transverse energy $E_T$ is to be expected
from the BFKL than from the DGLAP type of evolution in a
 rapidity interval between the proton fragmentation region 
 and the current fragmentation region.
A similar expectation holds for the inclusive transverse momentum distribution of 
hadrons~\cite{kuhlen,fbw:paper}.
So far, however, none of the studied observables (including recent measurements of
forward jet and large-$p_T$ charged and neutral 
pions~\cite{forward:jet})
are directly probing the correlation structure of the parton cascade.
To allow additional discrimination between a strictly $k_T$-ordered and
an unordered scenario, it is necessary to go beyond
single-particle observables and
consider directly measures of the transverse momentum correlations in
the parton cascade.

\section{The transverse momentum transfer}

In this paper we consider the 
transverse momentum transfer across rapidity $y$, $\vec{\Pi}(y)$.
This quantity is defined as the vector-sum of the transverse momenta $\vec{k_i}$ 
of all particles with
rapidities smaller than $y$ in an event with $n$ particles. 
For simplicity, we consider in the following only one of the components 
of the transverse momentum vectors (denoted by $k_i$) and define
\begin{equation}
\PI(y)=\sum_{i=1}^n k_i\,\theta(y-y_i),\label{eq:1}
\end{equation}
where $\theta(x)=0$ for $x<0$ and $\theta(x)=1$ for $x\geq0$.  
$\PI(y)$ is a random function, varying  from event to event. A set of 
$k$-th order moment-functions
can be constructed defined as\footnote{See~\cite{stratonovich} for 
a discussion of the theory of random functions}
\begin{equation}
\av{\PI(y_1)\,\PI(y_2)\ldots\PI(y_k)},\label{eq:moments}
\end{equation}
where $\av{}$ means that the average is taken over an ensemble of events.
From the moment-functions, one can further  define a set of correlation functions by 
taking a cluster decomposition (see eg.~\cite{review:93}). 
The two-point correlation function
$D^2(y_1,y_2)$ is then given by
\begin{equation}
D^2(y_1,y_2)=
\av{\PI(y_1)\,\PI(y_2)}- \av{\PI(y_1)}\,\av{\PI(y_2)}.\label{eq:corr}
\end{equation}
In particular, for $y_1=y_2=y$ we define $D^2(y)=D^2(y,y)$. This quantity measures the
variance  of the transverse momentum transfer distribution 
across the rapidity ``boundary'' $y$. 
$D^2(y)$ provides a measure of the rapidity structure of the correlations between the
transverse momenta as can be seen from the relation~\cite{bialas:1975}
\begin{equation}
D^2(y)=-\int_{-\infty}^y\,dy_1\int_{y}^{\infty}\,dy_2\int \,d^2k_1\,d^2k_2\,
\vec{k}_1\cdot \vec{k}_2\,\,\rho_2(\vec{k}_1,y_1;\vec{k}_2,y_2),\label{eq:d2}
\end{equation}
where $\rho_2(\vec{k}_1,y_1;\vec{k}_2,y_2)$ is the inclusive two-particle density.

With reference  to the parton-level diagram of Fig.~\ref{fig:ladder}, 
it should be noted that $\PI(y)$ is in that case equal to
the (propagator) transverse momentum exchanged between the vertices $i$ and $i+1$ 
if the position $y$ is located between particles $i$ and $i+1$ on the rapidity axis. 
Transverse momentum transfer correlations are, therefore, 
a direct measure of the correlations between the exchanges.


In models of uncorrelated production, the transverse momentum of a given
particle can be compensated by that of any other particle or group of particles.
Consequently, the correlation length describing
the transverse momentum correlations, which follow from momentum conservation
alone, is of the order of half the available rapidity
range and should thus increase with energy or $W$ in DIS.
For uncorrelated transverse momenta $D^2(y)$ is expected to increase when
$y$ moves from the edge of the rapidity space to $y=0$ since, for independent
random variables, the variance of a sum increases proportionally to the number
of terms included.  For the same reason also $D^2(0)$ should increase with the
number of particles produced.

Going beyond the uncorrelated production assumption, one 
could assume that the relevant
part of the diagram in Fig.~\ref{fig:ladder} 
(neglecting any dependence on longitudinal
momenta) can be schematically written as a product 
of nearest-neighbour correlated ``links''
\begin{equation}
|T_n|^2\approx \prod_i \,\, V(\vec{\PI}_i,\vec{\PI}_{i+1}),\label{eq:links}
\end{equation}
where the functions $V$ represent the (iterated) kernel of the diagram.
In that case, an important role is played by the eigenvalues and eigenfunctions of the
kernel, considered as a transfer integral operator~\cite{michael}.
For a sufficiently long ladder, it can be shown that the function $D^2(y_1,y_2)$
is controlled by the largest eigenvalue $\lambda_{01}$ leading to a behaviour
\begin{equation}
D^2(y_1,y_2)\approx 
\exp\left\{
-\rho(1-\lambda_{01})|y_1-y_2|
\right\},
\label{eq:d2eigen}
\end{equation}
with $\rho$ the mean parton multiplicity per unit of rapidity.
In this short-range model, the correlation length is therefore given by
$1/\rho(1-\lambda_{01})$.

In models with short-range correlations, by definition, particles far away in
rapidity cannot be correlated. As a result, the compensation of transverse
momenta is local, i.e.~it is compensated by its neighbours. The corresponding
correlation length $\lambda$ is energy independent.
These general properties are reflected in the behaviour of $D^2(y)$.
Only particles in the rapidity interval $\Delta=(-\lambda+y,y+\lambda)$
contribute to the transverse momentum transfer at $y$. 
Therefore, $D^2(y)$ is expected  not to increase faster
with energy than the particle density in the interval $\Delta$
(for constant $\lambda$) which is known to increase only  logarithmically.

In a $k_T$-ordered cascade, correlations will be long-range. The correlation length
can be expected to increase with the `length of the cascade' in rapidity space,
i.e.~proportional to $\ln{1/x}$ or $\ln{W}$ (at fixed $Q^2$) in DIS.
For the case of the BFKL scenario, it can further be expected that the correlation
length is related to the  dominant eigenvalues of the BFKL kernel
(cfr.\@~Eq.~\ref{eq:d2eigen}) 

\section{A Monte Carlo study}

Predictions for the cases of the
ordered and unordered cascades are obtained
from DIS Monte Carlo models, which incorporate the QCD evolution in
different approximations and utilize phenomenological models
for the non-perturbative hadronization phase.
The MEPS model (Matrix Element plus Parton Shower)~\cite{bib:lepto65},
incorporates the QCD matrix elements up to first order, with additional
soft emissions generated by adding leading log parton showers.
In the colour dipole model (CDM)~\cite{dipole,bib:ariadne}
radiation stems from colour dipoles formed by
the colour charges.  Both programs
use the Lund string model~\cite{bib:jetset} for hadronization.
The CDM description
of gluon emission is similar to that of the BFKL evolution,
because the gluons emitted by the dipoles
do not obey strong \kt ordering~\cite{bfklcdm}. In MEPS the partons
are strongly ordered in $k_T$, because they are based upon
leading log DGLAP parton showers.

Two samples of one million events have been generated using the latest versions of the models
(Lepto 6.5 for MEPS, Ariadne 4.08 for CDM) with the parton density parametrisation GRV-94~\cite{GRV}.
The fraction of events undergoing a Soft Colour Interaction~\cite{sci:ingelman} was set to zero.
In all figures the transverse momentum transfer was calculated using final state hadrons,
including neutrals so that transverse momentum is conserved for the full event.

\begin{figure}
\centering
\epsfig{file=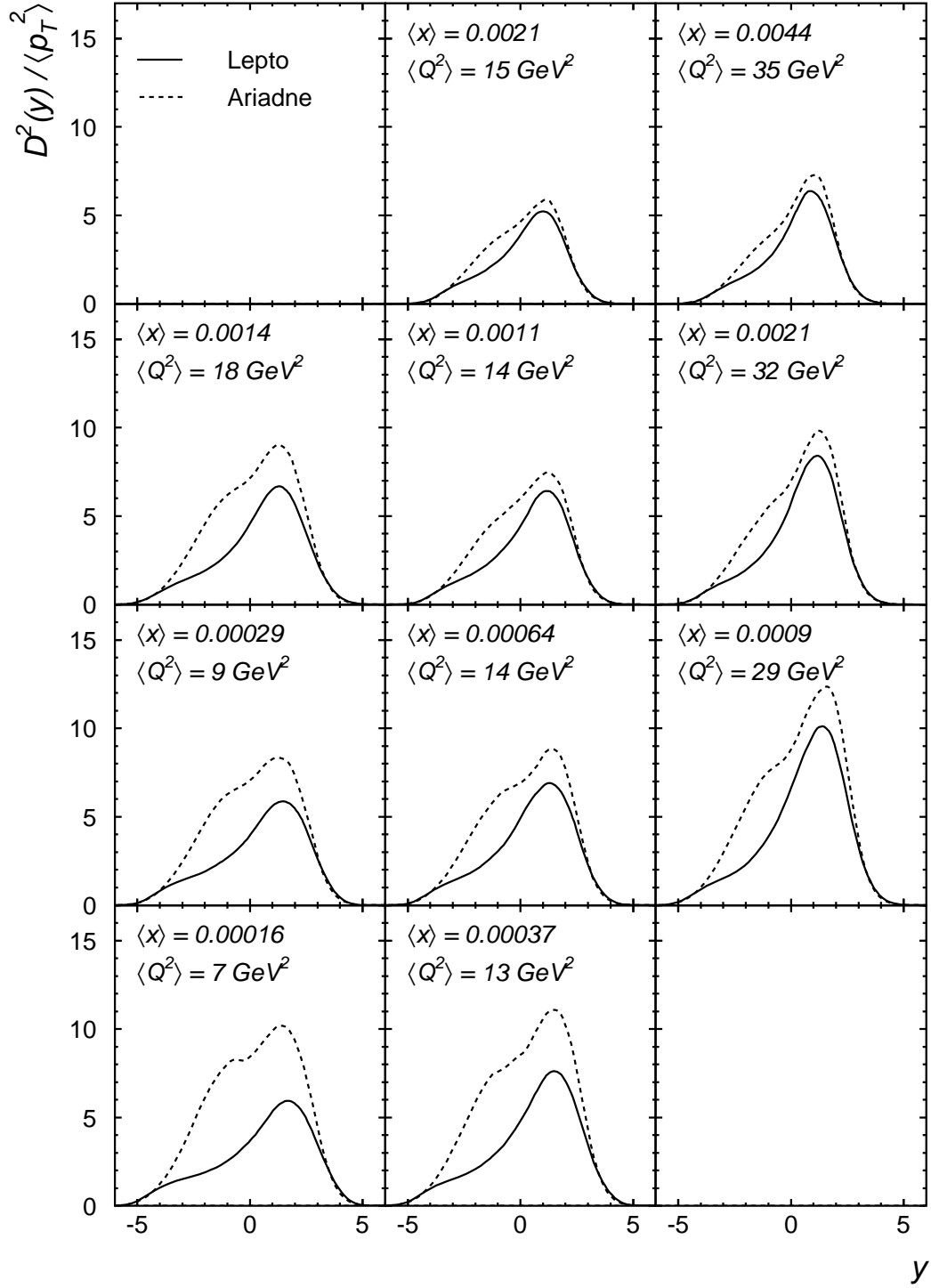,height=20cm}
\caption{The fluctuation of the transverse momentum transfer, $D^2$,
scaled to the average transverse momentum squared, $\langle p_T^2 \rangle$, is shown as
a function of rapidity $y$ in different bins of $Q^2$ and $x$ for Lepto (full line) and Ariadne
(dashed line).}
\label{fig:dispersion}
\end{figure}

Figure~\ref{fig:dispersion} shows the variance of the transverse momentum transfer, $D^2(y)$,
scaled to the average hadron transverse momentum squared in different bins of $Q^2$ and $x$.
Large differences between models
are observed at low $x$.  Because the variance is scaled, this effect is not due to the larger
$p_T$-flow produced by Ariadne.  As expected, the transverse momentum transfer fluctuations
are larger in the unordered scenario.

\begin{figure}
\centering
\epsfig{file=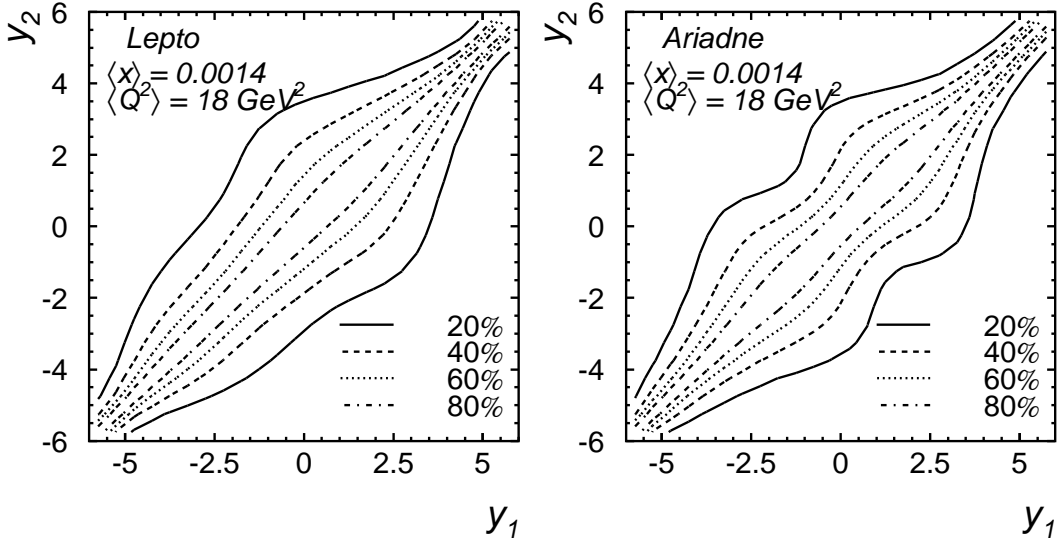,height=10cm}
\caption{The two-dimensional correlation function, defined as the ratio $D^2(y_1,y_2) / D(y_1) D(y_2)$,
is shown as a contour graph for Lepto (top) and Ariadne (bottom).}
\label{fig:corr_matrix}
\end{figure}

\begin{figure}
\centering
\epsfig{file=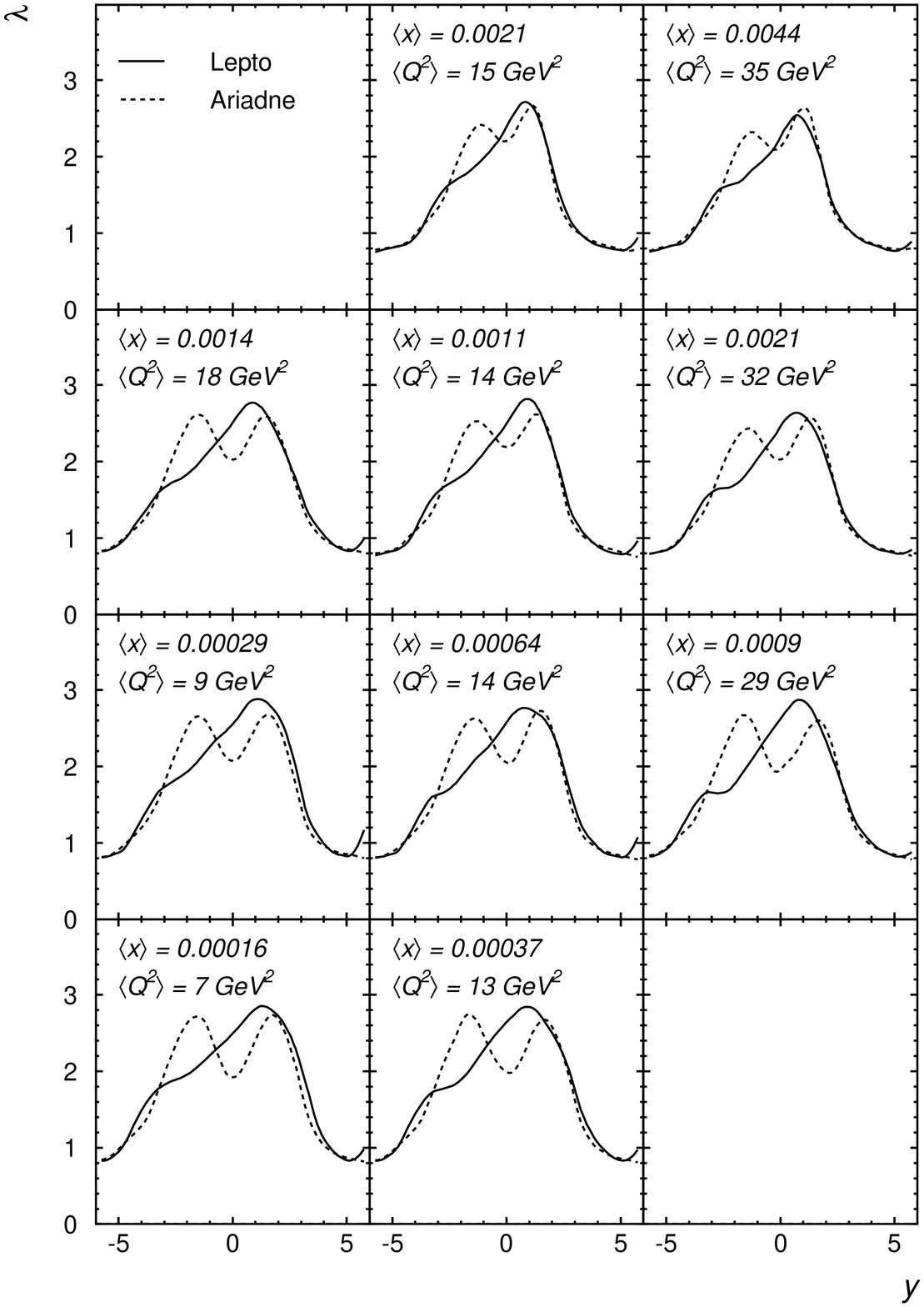,height=20cm}
\caption{The correlation length, $\lambda$, defined as in the text, is shown as a function of rapidity
$y$ in different bins of $Q^2$ and $x$ for Lepto (full line) and Ariadne (dashed line).}
\label{fig:corr_length}
\end{figure}

The correlation function, computed as the ratio $D^2(y_1,y_2) / D(y_1) D(y_2)$, is shown in
Fig.\@~\ref{fig:corr_matrix}.  The correlation length (shown in Fig.\@~\ref{fig:corr_length}) is calculated along slices perpendicular to
the diagonal $y_1 = y_2$.  Within these slices the correlation decreases exponentially with $|y_1 - y_2|$
and the correlation length is defined as the slope of an expontial fit.  The models used in this
simulation show no strong dependence of the correlation length on $x$ or $Q^2$.  Again, differences
are observed in the behaviour of the correlation length between the Lepto and Ariadne Monte Carlo models.  In particular, Ariadne predicts a symmetric behaviour in the proton and photon fragmentation hemisphere with a double-peaked structure for the correlation length, while Lepto predicts longer correlation lengths in the photon fragmentation hemisphere.

\section{Conclusion}

It has been demonstrated that the transverse momentum transfer correlation is a theoretically attractive
variable to discriminate between models for QCD evolution in parton cascades.  The two considered models,
Lepton and Ariadne, show large differences in the behaviour of the transverse momentum transfer fluctuation
and correlation length.  However, further work is needed to investigate the feasibility of an experimental
measurement.

\end{document}